# Rainbows in a bottle: Realizing microoptic effects by polymerizable multiple emulsion particle design

Naresh Yandrapalli[ab*], Baris Kumru[a], Tom Robinson[b*], Markus Antonietti[a]

[a] Max Planck Institute of Colloids and Interfaces, Department of Colloid Chemistry, Am Mühlenberg 1, 14424 Potsdam, Germany

[b] Max Planck Institute of Colloids and Interfaces, Department of Theory & Bio-Systems, Am Mühlenberg 1, 14424 Potsdam, Germany

## Introduction

In nature, structural colour generation is based on discriminative light propagation associated with physical structures in the range of the wavelengths of light[1]. These iridescent structural colours are of immense significance[2] but not easy to control experimentally and therefore difficult to exploit for applications. In this work, we employ microfluidics to produce polymerizable double emulsions that can not only induce the already known lensing effect[3] but also result in the spectral separation of white light. Here, liquids of varying refractive index that constitute the emulsions resulted in patterns of iridescent colours. After polymerization, the inner emulsion cores collapse and this results in curved concave surfaces on these polymeric microspheres. Interestingly, the light propagation along the curved surfaces undergo total internal reflection, followed by near-field interference along exit structures on the polymerized microspheres[4]. These structured polymeric particles that are able to generate colour dispersions can be exploited for optical devices, displays and even sensing technologies.

## Main

From the formation of glories in the sky to the spectacular vibrant colours observable on various living organisms, humans have learned that light can be controlled by materials structures/structured materials that exist and evolved in nature. We are continuously discovering that the majority of the everyday iridescent spectra, either colourful butterflies or the plumage of birds, is the result of micrometre scale material structures[5,6]. Simple physical phenomena such as absorption, reflection, and refraction, as well as diffraction and interference that result from light-matter interactions can combine into complex structural colour generations [5]. An attempt to classify the physical interaction of light with microstructures resulted in identifying processes such as thin-film interference[7], multi-film interference[8], diffraction grating[9], scattering (coherent & incoherent)[10] and photonic crystal diffraction[11] that leads to structural colouration observed in nature[12]. Recently, Goodling et al. have shown that light interaction with concave interfaces formed by materials with two different refractive indices ($\eta$) will result in structural colouration due to the interference of total internally reflected rays[4]. This proves that two materials (with varying refractive indices) forming a simple interface with a degree of curvature can produce iridescent colours.

Our results show that water-in-oil-in-water (W/O/W) double emulsions produced with a highly refractive oil layer (styrene ($\eta$-1.516) interfaced with water ($\eta$-1.333) on either side can generate iridescent colours with spectral separation similar to that observed in glories[13] but with more complexity and can still be exploited for their light focusing effect[14]. The double emulsions produced here contain an aqueous inner core surrounded by a layer of styrene in a continuous aqueous phase (shown in figure 1) stabilized with 0.5 wt% F108 surfactant. Because the two liquids have different refractive indices, light undergoes both refraction and reflection based on the layering of the two liquids. The spectral dispersions or rainbows observed in nature are an effect of light interacting with water drops in the air, however, in the case of double emulsions (here, W/O/W), having layers of aqueous-oil-aqueous-oil-aqueous phase result in a similar dispersion but are followed by merging of the spectral wavelengths to produce structural colouring (figure 1(b)). A z-stack of the diffused structural

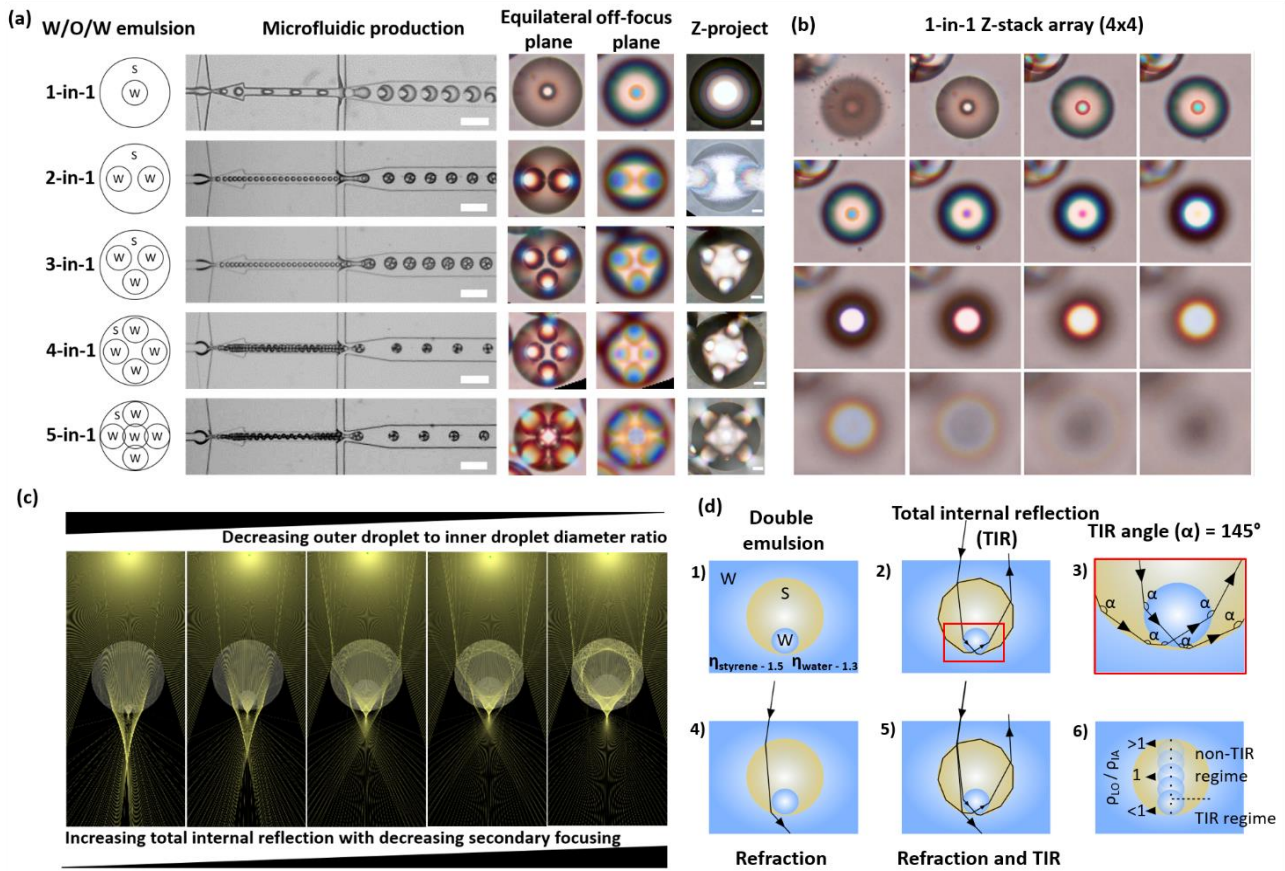

**Figure1:** Microfluidic generation of liquid emulsions. (a) Double and multi-core emulsion production with varying inner cores (scale bar – 200 µm) and the iridescent nature of the dispersed light at the equatorial plane, off-focus plane (30 µm below the equatorial plane) of the emulsions and the intense light spots observed along the z-axis of the inner cores (scale bar – 10 µm). (b) bright-field z-stacks of the light scattering from a 1-in-1 emulsion with each slice separated by 10 µm (top left to bottom right). (c) 2D ray-tracing of a 1-in-1 emulsion with varying inner droplet sizes (diameters 5, 10, 30, 40 and 50 µm) showing the effect on the focal length and TIR. (d) schematic representation of a 1-in-1 emulsion showing refraction, TIR and the effect of inner to outer droplet density on the TIR (black arrows represent the direction of light with a single wavelength). S – corresponds to styrene phase and W – corresponds to aqueous phase.

colouring produce from a 1-in-1 W/O/W double emulsion is presented in supplementary video 1, and a clear spectral separation is modelled by ray tracing (see supplementary video 2). As depicted schematically in figure 1d(4), the light rays from the source passing through the double emulsion initially undergo refraction at the first convex interface formed by the outer aqueous solution and the oil layer as well as through the second convex interface across the aqueous inner core present inside the oil drop. Thanks to the spherical configuration of the emulsion, a single light ray can undergo a minimum of two or a maximum of four refraction events as it passes out of the emulsion (1d). This results in more complex colour formations than observed from single emulsions.

Our observations suggest that the focusing effect from these double emulsions is also prominent (see figure 1c), which was not observed in previous works on diverging light rays passing through low refractive index inner core surrounded by a high refractive index liquid[3]. Moreover, a point source of light passing through the emulsions such as the present case resulted in dual focusing – one right below the double emulsion (resulting from the rays passing through the inner aqueous core) and the other farther away (resulting from the rays passing through the rest of the emulsion) (see supplementary figure S2). In agreement with that, z-stack images of the double emulsion indeed show a bright spot right below the emulsion due to the said focusing of the dispersed light beyond the emulsion (see supplementary video 1 as well as z-project from figure 1(a)). The second focus vanishes for the double emulsions whose inner aqueous core diameter is equal to or above the radius of the diameter of the W/O/W emulsion (see figure 1c). This is because the rays that are otherwise contributing to the second focus are then refracted through the bigger inner aqueous core converge into the primary focusing area. This is observed by the increase in the intensity of the primary focusing area as the inner aqueous droplet size increases (figure 1c).

Previous studies have shown that refraction through equidistant alternating layers of liquid or transparent material is associated with multi-layer interference[8]. This phenomenon is not observable in the case of the double emulsions described here. The inner aqueous droplet being heavier ($\rho$ = 0.9982 g/mL at 20 ºC) than the oil phase ($\rho$ = 0.909 g/mL at 20 ºC) sinks to the bottom of the oil drop (figure 1d(1)), resulting in non-equidistant layers and a decrease in peak reflectivity[8]. As the light is passing through a high refractive index medium to a lower refractive index medium, followed by the curved nature of the interface[4], a total internal reflection (TIR) of light is possible at particular angular incidences, observable from the 2D ray-tracing result (figure 1c) and presented in the scheme (figure 1d(2)). Only refracted rays whose angle of incidence ($\alpha$) is ~145° can undergo TIR when they pass through the inner aqueous core and encounter the concave interface across styrene (high refractive index) and the aqueous outer solution (low refractive index) (figure 1d(3)). The schematic produced with the ray tracing data figure 1d(6) shows the density-dependent TIR with $\rho_{styrene}/\rho_{water}$ less than 1, while values greater than 1 result in no TIR. This suggests that a double emulsion with a high refractive index outer layer and a low refractive index inner layer should require inversed density values to take advantage of TIR as shown previously.[4]

Similarl to double emulsions, the emulsions with multiple inner droplets inside one W/O/W double emulsion generated by the same single inlet microfluidic device (see Methods) indeed showed light scattering and iridescent colours (figure 1a). Producing such multi-core emulsions using a single inlet microfluidic design was made possible in this work by taking advantage of the swelling property of PDMS in the presence of specific solvents like styrene.[15,16] The swelling of PDMS upon the uptake of styrene resulted in narrowed channels width, especially at first cross junction where water-in-oil (W/O) droplets are formed. Through carefull manipulation of flow rates, when can achieve W/O/W emulsions wih multiple inner cores, all with the same device design (see figure 1a and Supplementary Video 3). In the absence of such a reliable system, multiple device designs with varied channel widths would have to be fabricated. The light scattering observed in all these multi-core emulsions, 2-in-1, 3-in-1, 4-in-1, and 5-in-1, are presented in the supplementary figure S1 as arrays of z-stack slices and in supplementary video 4, 5, and 6. At any given z-slice, the light scattering observed is similar for each different type of multi-core emulsion. There is no observable influence of the light scattering produced by one inner droplet on the other inner droplet within such multi-core emulsions. Moreover, the iridescence within and around each inner droplet is preserved. This is only possible if the inner droplets are in the same plane (at the bottom) and are not stacked one above the other (see ray tracing data in supplementary figure S2). Although the formation of emulsions with multiple inner droplets has been presented previously, their light scattering and high iridescence properties have never been explored before[17,18].

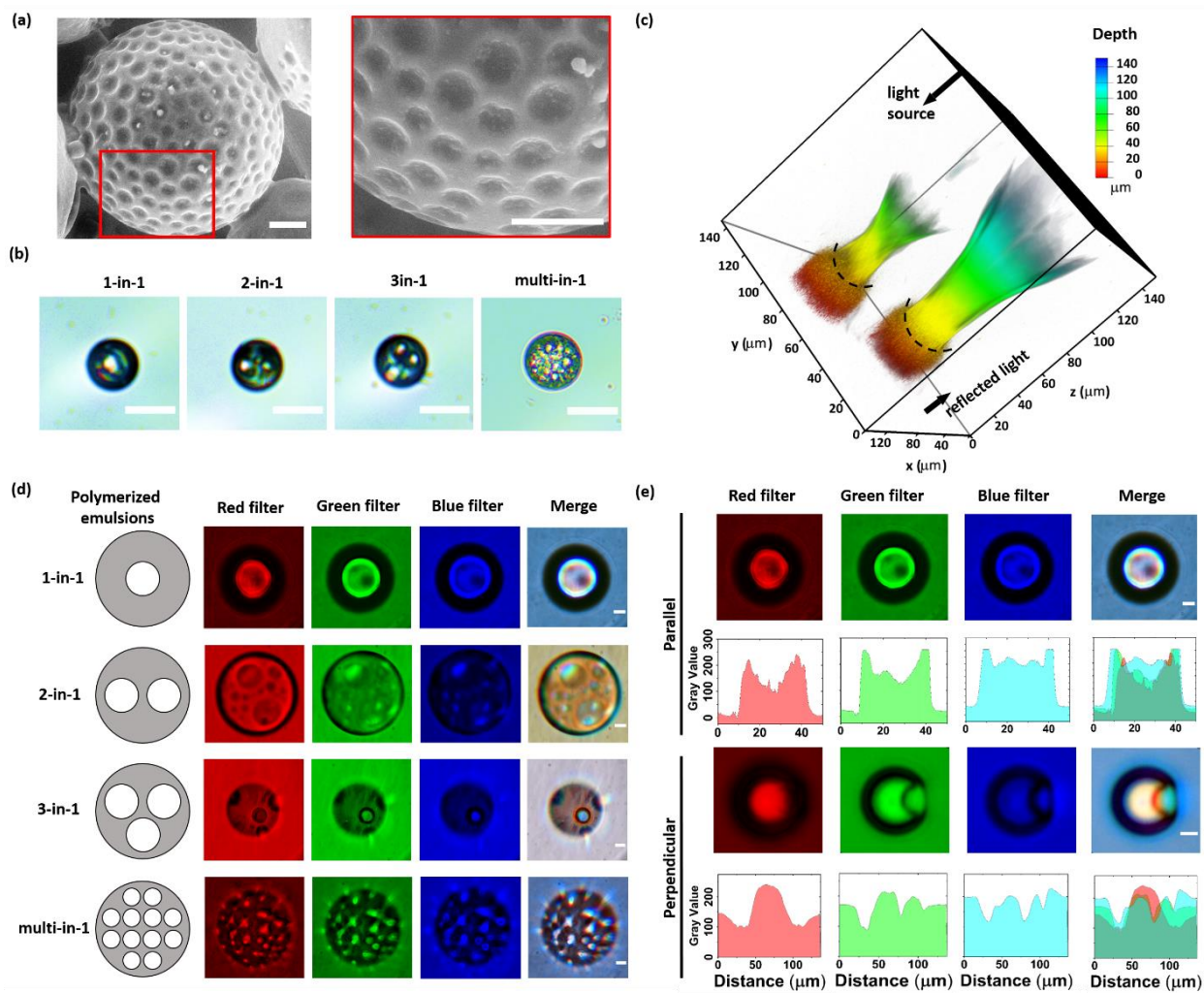

**Figure 2**: Light scattering from polymerized emulsions. (a) Electron micrographs of polymerized multi-core emulsion, inset showing the close-up of the curved surfaces on the microsphere (scale bar – 10 μm). (b) Bright-field images of the polymerized emulsions with a varying number of inner cores (scale bar – 50 μm). (c) Confocal 3D reconstruction of the reflected light from the curved surfaces of two polymerized 1-in-1 emulsions facing the light source. The curved dashed lines represent the surface boundary of the concave surface on the microparticle . (d) Color images of the polymerized multi-core emulsions showing spectral dispersion of the reflected light from the curved surfaces (scale bar – 10 μm). (e) Spectral separation of dispersed reflected light

from the curved surfaces of polymerized 1-in-1 emulsion aligned 0° (parallel) and 90° (perpendicular) angle to the light source (scale bar – 10 µm).

We then further explored the possibility to polymerize the styrene within the double and multi-core emulsions, with the aim of preserving some of these effects into a more stable, polymer structure. We have demonstrated that microfluidics can be used to generate uniform sized hollow microspheres [19]. However, it is not possible to produce structural colouration/iridescent colours with individual microspheres or hollow microspheres unless they are transparent or with surface patterning[20] or regularily packed to form so-called photonic balls[21].

In this work, the above produced double emulsions and multi-core emulsions with styrene as the oil-phase are polymerized to generate microspheres with lensing curved surfaces. Unlike for the formation of hollow polymer microspheres where the inner aqueous cores have to be stabilized, microspheres with concaved dimples as shown in figure 2 are made from inner aqueous cores devoid of any surfactant. The surfactant F108 used in this study is a tri-block copolymer with two polypropylene hydrophobic blocks separated by a central hydrophilic polyethene oxide block, which is known to form rather stable block copolymer bilayers along two aqueous droplets, similar to that of lipids in a cell membrane[22]. However, in the absence of surfactant in the inner aqueous core, there will only be a monolayer of the surfactant stabilizing the emulsion interface. We hypothesize that during polymerization, destabilization of the monlayer could result in the formation of a concaved dimple on the spherical polystyrene microsphere as the inner aqueous core fuses with the outer aqueous solution (as shown schematically in supplementary figure S3). The scanning electron micrographs of these polymeric spheres show the resulting concave surfaces (see figure 2(a)). Similarly, polymerization of 2-in-1, 3-in-1, and multi-in-1 emulsions resulted in an equivalent number of surface lenses on the respective polystyrene microspheres formed (see bright-field images in figure 2(b)).

Unlike the averaged light scattering observed in liquid phase emulsions, single polystyrene microspheres can be described via the rules of the reflection of light[23] and thereby show optical near field effects. Like concave mirrors, the dimples on the polymeric microsphere reflect and focus the light. True to this assumption, the 3D reconstruction of the confocal z-stack images suggests both reflection and focused light from the structured surface of the polymerized 1-in-1 W/O/W emulsion (figure 2(c)). The z-stack was acquired with the curved surface facing towards the source of light.

Similar to the liquid emulsions, our observations of their polymerized versions using a colour camera have revealed the iridescence nature of these focused surface emissions. Initial observations of these particles have shown their surface dimples/curves and iridescent colouring around the particles as well as bright spots from the lensing regions (figure 2(b)). A closer look reveals more spectacular details of iridescence from the lensing regions of the particles (figure 2(d)). Concaved structures that are facing the light source reflect the light and appear as intense bright white spots – a combination of all the spectral colours, as seen in figure 2(b) and 1-in-1 polymerized emulsions of figure 2(d). Interestingly, particles with concaved structures which are not facing but at an angle to the light source show a dramatic colour separation. This is exhibited in figure 2(d) for 2-in-1, 3-in-1 and multi-in-1 configurations of the polymerized emulsions. In the case of multi-in-1 particles, with multiple curved surfaces facing the light source at different angles (figure 2(d)), one can observe that approximately 20 micron "rainbows" show no apparent angular dependency, within the observed upper hemisphere of the particle . Light reflected from the concave interface of polymeric microspheres creates optical interference and the dispersion of light[4]. In our experiments with the results shown in figure 2, this is clearly visible that the longer wavelengths of the reflected light are closer to the surface of the microsphere while rays of shorter wavelengths are farther away from the surface (see figure 2(e)). The line profile data taken from the centre of the images plotted to suggest the mixing and separation of the three main wavelengths - red, green and blue. A similar patterning of light reflection and dispersion is observed in polymeric microspheres, 2-in-1, 3-in-1, and multi-in-1 (figure 2(d)). This suggests that not only the liquid emulsions but also their polymerized versions can be exploited for light scattering properties.

This study highlights some interesting optical phenoma that are enabled by manufacturing controlled double and multi-core emulsions with a high refractive index differences using microfluidics. We have shown that simple emulsions like W/O/W can induce spectral separation of white light under very local light focusing. The observed light enhancements are mainly due to refraction and to a lesser extent from total internal reflection. More dramatic colour patterns can be achieved by simple alteration of the double emulsions to complex multi-core emulsions, also involving morphological transitions occurring under polymerization. While structural colouring in ordered 2D-arrays is a clear use case as a multilens array, because of the focusing nature, the applications and use of the local, close-to particle light fields with their colour gradients are yet to be discovered. The possibility to embed catalysts within the polymerizable liquid emulsions was demonstrated earlier[19] and opens applications towards focused photocatalysis and photocatalytic gradients. Well ahead of applicability and exploitation in multiple fields however, we want to underline that the particles presented above are simple and rather effective to make, and that the observed effects are also just stunningly beautiful and unexpected: little rainbows in a bottle.

# References

1. Sun J, Bhushan B, Tong J. Structural coloration in nature. RSC Adv. 2013; **3**: 14862–14889.
2. Cuthill IC, Allen WL, Arbuckle K, Caspers B, Chaplin G, Hauber ME *et al.* The biology of color. Science (80-. ). 2017; **357**. doi:10.1126/science.aan0221.
3. Nagelberg S, Zarzar LD, Nicolas N, Subramanian K, Kalow JA, Sresht V *et al.* Reconfigurable and responsive droplet-based compound micro-lenses. *Nat Commun* 2017. doi:10.1038/ncomms14673.
4. Goodling AE, Nagelberg S, Kaehr B, Meredith CH, Cheon SI, Saunders AP *et al.* Colouration by total internal reflection and interference at microscale concave interfaces. *Nature* 2019; **566**: 523–527.
5. Sun J, Bhushan B, Tong J. Structural coloration in nature. RSC Adv. 2013; **3**: 14862–14889.
6. Saranathan V, Narayanan S, Sandy A, Dufresne ER, Prum RO. Evolution of single gyroid photonic crystals in bird feathers. *Proc Natl Acad Sci U S A* 2021; **118**: 2101357118.
7. Kolle M, Salgard-Cunha PM, Scherer MRJ, Huang F, Vukusic P, Mahajan S *et al.* Mimicking the colourful wing scale structure of the Papilio blumei butterfly. *Nat Nanotechnol* 2010; **5**: 511–515.
8. Kinoshita S, Yoshioka S, Miyazaki J. Physics of structural colors. *Reports Prog Phys* 2008; **71**: 076401.
9. Parker AR. Natural photonics for industrial inspiration. *Philos Trans R Soc A Math Phys Eng Sci* 2009; **367**: 1759–1782.
10. Prum RO. A Fourier Tool for the Analysis of Coherent Light Scattering by Bio-Optical Nanostructures. *Integr Comp Biol* 2003; **43**: 591–602.
11. Joannopoulos JD, Johnson SG, Winn JN, Meade RD. *Photonic crystals: Molding the flow of light*. 2011http://ab-initio.mit.edu/book/ (accessed 12 Feb2021).
12. Parker AR. 515 million years of structural colour. J. Opt. A Pure Appl. Opt. 2000; **2**. doi:10.1088/1464-4258/2/6/201.
13. Laven P. Atmospheric glories: Simulations and observations. In: *Applied Optics*. Optical Society of America, 2005, pp 5667–5674.
14. Calixto S, Rosete-Aguilar M, Sanchez-Marin FJ, Marañon V, Arauz-Lara JL, Olivares DM *et al.* Optofluidic compound microlenses made by emulsion techniques. *Opt Express* 2010. doi:10.1364/oe.18.018703.
15. Lee JN, Park C, Whitesides GM. Solvent Compatibility of Poly(dimethylsiloxane)-Based Microfluidic Devices. *Anal Chem* 2003. doi:10.1021/ac0346712.
16. Dangla R, Gallaire F, Baroud CN. Microchannel deformations due to solvent-induced PDMS swelling. *Lab Chip* 2010. doi:10.1039/c003504a.
17. Wang W, Xie R, Ju XJ, Luo T, Liu L, Weitz DA *et al.* Controllable microfluidic production of multicomponent multiple emulsions. *Lab Chip* 2011. doi:10.1039/c1lc20065h.
18. Vladisavljević GT, Al Nuumani R, Nabavi SA. Microfluidic production of multiple emulsions. Micromachines. 2017. doi:10.3390/mi8030075.
19. Yandrapalli N, Robinson T, Antonietti M, Kumru B. Graphitic Carbon Nitride Stabilizers Meet Microfluidics: From Stable Emulsions to Photoinduced Synthesis of Hollow Polymer Spheres. *Small* 2020; **16**: 2001180.
20. Yeo SJ, Park KJ, Guo K, Yoo PJ, Lee S. Microfluidic Generation of Monodisperse and Photoreconfigurable Microspheres for Floral Iridescence–Inspired Structural Colorization. *Adv Mater* 2016; **28**: 5268–5275.
21. Vogel N, Utech S, England GT, Shirman T, Phillips KR, Koay N *et al.* Color from hierarchy: Diverse optical properties of micron-sized spherical colloidal assemblies.

*Proc Natl Acad Sci U S A* 2015. doi:10.1073/pnas.1506272112.
22   Kulkarni C V. Lipid crystallization: From self-assembly to hierarchical and biological ordering. *Nanoscale* 2012; **4**: 5779–5791.
23   Wang M, Ye X, Wan X, Liu Y, Xie X. Brilliant white polystyrene microsphere film as a diffuse back reflector for solar cells. *Mater Lett* 2015; **148**: 122–125.
24   Yandrapalli N, Petit J, Bäumchen O, Robinson T. Surfactant-free production of biomimetic giant unilamellar vesicles using PDMS-based microfluidics. *Commun Chem 2021 41* 2021; **4**: 1–10.
25   Tu R, Johnson. Ray Optics Simulation - Home. https://ricktu288.github.io/ray-optics/ (accessed 11 Feb2021).

## Supplementary information

Materials, supplementary Figs. 1-4 and caption for supplementary Video 1-6.

## Acknowledgments

The authors thank the Max Planck Society for funding. T.R. and N.Y. acknowledge support from the MaxSynBio consortium, which is jointly funded by the Federal Ministry of Education and Research of Germany and the Max Planck Society.

## Author Contributions

N.Y., B.K., T.R., and M.A. conceptualized the project. N.Y. performed the experiments. N.Y., T.R., and M.A. analyzed the data and wrote the manuscript.

## Author Information

Naresh Yandrapalli

Department of Colloid Chemistry, Max Planck Institute of Colloids and InterfacesAm Mühlenberg 1, Potsdam 14424, Germany

Email: naresh.yandrapalli@mpikg.mpg.de

Baris Kumru

Department of Colloid Chemistry, Max Planck Institute of Colloids and InterfacesAm Mühlenberg 1, Potsdam 14424, Germany

Email: baris.kumru@mpikg.mpg.de

Tom Robinson

Department of Theory & Bio-Systems, Max Planck Institute of Colloids and Interfaces, Am Mühlenberg 1, Potsdam 14424, Germany

E-mail: tom.robinson@mpikg.mpg.de

Markus Antonietti

Department of Colloid Chemistry, Max Planck Institute of Colloids and InterfacesAm Mühlenberg 1, Potsdam 14424, Germany

E-mail: markus.antonietti@mpikg.mpg.de

## Data Availability

All data generated or analysed during this study are included in the published article and supplementary information, and are available from the corresponding authors upon reasonable request.

## Competing Interest Declaration

No competing interests to declare.

## Corresponding Author

Correspondence to Naresh Yandrapalli and Tom Robinson

## Figure Legends

**Figure1:** Microfluidic generation of liquid emulsions. (a) Double and multi-core emulsion production with varying inner cores (scale bar – 200 µm) and the iridescent nature of the dispersed light at the equatorial plane, off-focus plane (30 µm below the equatorial plane) of the emulsions and the intense light spots observed along the z-axis of the inner cores (scale bar – 10 µm). (b) bright-field z-stacks of the light scattering from a 1-in-1 emulsion with each slice separated by 10 µm (top left to bottom right). (c) 2D ray-tracing of a 1-in-1 emulsion with varying inner droplet sizes (diameters 5, 10, 30, 40 and 50 µm) showing the effect on the focal length and TIR. (d) schematic representation of a 1-in-1 emulsion showing refraction, TIR and the effect of inner to outer droplet density on the TIR (black arrows represent the direction of light with a single wavelength). S – corresponds to styrene phase and W – corresponds to aqueous phase.

**Figure 2**: Light scattering from polymerized emulsions. (a) Electron micrographs of polymerized multi-core emulsion, inset showing the close-up of the curved surfaces on the microsphere (scale bar – 10 µm). (b) Bright-field images of the polymerized emulsions with a varying number of inner cores (scale bar – 50 µm). (c) Confocal 3D reconstruction of the reflected light from the curved surfaces of two polymerized 1-in-1 emulsions

facing the light source. The curved dashed lines represent the surface boundary of the concave surface on the microparticle . (d) Color images of the polymerized multi-core emulsions showing spectral dispersion of the reflected light from the curved surfaces (scale bar – 10 µm). (e) Spectral separation of dispersed reflected light from the curved surfaces of polymerized 1-in-1 emulsion aligned 0° (parallel) and 90° (perpendicular) angle to the light source (scale bar – 10 µm).

## Methods

### Device Fabrication

Master mould for the microfluidic device was created using UV-based photolithography. Initially, a 4 inch silicon wafer was pre-heated for 30 min at 200 °C and 80 µm thick layer of photoresist (SU8 2025) was spin-coated on top (model no. WS-650MZ-23NPPB, Laurell Tech. Corp) as per the specifications provided by the manufacturer at 23 °C. After the coating step, the wafer is heated at 65 °C for 3 min and 95 °C for 9 min before UV exposure. After 8 sec of UV exposure through a specific design (see the supplementary figure S4) using kloe photolithographic instrument (model no. UV-KUB 2), the wafer was post-baked at 65 °C for 2 min and 95 °C for 7 min. The device design was revealed on the wafer after the washing steps with a developer solution and isopropanol. Finally, the wafer was baked at 200 °C for 2 h before performing overnight surface passivation with 50 µL of 1H,1H,2H,2H-perfluorodecyltrichlorosilane in a dessicator.

To produce the microfluidic chips, a PDMS:curing agent (10:1) mixture was thoroughly mixed and degassed for 30 min in a desiccator connected to low pressure (150 millibars). The degassed mixture was poured on top of the surface passivated wafer and cured at 90 °C for 3 h. Crosslinked PDMS was pealed from the master mould and diced into small pieces. The inlets and outlets were created using 1 mm biopsy puncher (Kai Europe GmbH). Finally, plasma cleaned (at 600 mbar for 1 min) (Plasma Cleaner PDC-002-CE, Harrick Plasma) glass coverslips and diced PDMS chips with the desired design were bonded to form the microfluidic chip. These chips were further heated at 60 °C for 2h to complete the bonding process and retention of hydrophobic surface.

### Device Surface Passivation

Surface passivation of the double emulsion microfluidic design is necessary for the formation of stable double emulsions[24]. A series of solutions are flown through the outlet to the outer aqueous (OA) solution inlet to render the hydrophobic PDMS surface hydrophilic (see the supplementary figure S4(a)). To achieve this, initially, a 2:1 mixture of $H_2O_2$-HCl solution was flushed for 30 sec. This was followed by flushing of 10 wt% of PDADMAC solution for 2 min and later by 5 wt% of PSS solution for another 2 min. After every step, MilliQ® water was flushed for 30 sec to remove excess material. Thus, flushed solutions form a hydrophilic polyelectrolyte layer on top of the hydrophobic PDMS surface along the OA solution inlet to the outlet.

Production of Double and Multi-core emulsions

Briefly, the inner aqueous solution (IA) is flushed through the first cross-junction to form a water-in-oil (W/O) emulsion, followed by a second shearing step at the second cross-junction. This results in the formation of water-in-oil-in-water (W/O/W) double emulsion with a single inner aqueous core surrounded by styrene (S) which is stabilized through F108 (0.5 wt%) containing outer aqueous solution (OA) (see the supplementary figure S4(b)). Furthermore, to form multi-core emulsions, the swelling property of the PDMS in the presence of styrene is exploited to reproducibly narrow the channels at the first junction (at its lowest dimension) where water-in-oil (W/O) droplets are created. A careful alteration of the flow pressures resulted in controlling the number of aqueous droplets that get encased inside the final double emulsion. Using this property, double emulsions with double, triple, quadruple, and quintuple cores are produced. Since styrene is used as the oil phase, thus produced droplets can be polymerized to yield polymeric styrene microspheres. For fluid flow control, four-channel pressure devices are used (MFCS-EZ, Fluigent Inc.).

Emulsion polymerization

Double and multi-core emulsions prepared with styrene:octanol (95:5 %) oil phase dissolved with 0.5 wt% BAPO, were placed under UV illumination (395-400 nm, custom made device, 50W LED chips (Foxpic High Power 50 W LED Chip Bulb Light DIY White 3800LM 6500 K) and 30 W UV chip (Fdit, 395-400 nm UV LED chip) were connected to a self-made circuit and cooling system.) for 4 hours for complete polymerization of styrene. Produced microparticles were washed *via* centirifugation before imaging. Freeze drying was performed before scanning electron microscopy analysis.

Microscopy

Microfluidic production of multi-core emulsions is recorded with a MicroLab 310 camera at full-frame and ~3000 fps (Vision Research Inc.) that is connected to wide-field Olympus IX73 microscope using a x5 objective in bright-field transmission mode. Monochrome confocal images are acquired with Leica TCS SP8 (Leica Microsystems Inc.) confocal microscope. Colour images with Nikon DS-Fi3 high definition camera fitted to a wide-field Olympus IX73 microscope using a x40 objective in bright-field transmission mode. Additionally, red (580/30 nm), green (510/30 nm) and blue (420/30 nm) filters are also used to image the spectral separation. In all the cases, the 50 µL of emulsion suspensions were pipetted on to a 0.17 mm glass coverslip fitted with imaging spacers (SecurteSeal$^{TM}$) for imaging. General image processing is performed using ImageJ/Fiji, z-stack arrays with Huygens Professional (Scientific Volume Imaging Inc.), and 3D rendering of confocal z-stacks with LAS X Core (Leica) module.

Ray tracing

Ray tracing simulations are performed using Ray-Optics Simulation software[25] and COMSOL Multiphysics® (COMSOL AB, Stockholm, Sweden). Dimensions of the droplets and emulsions simulated were obtained from the microscopic images of the emulsions. Furthermore, the COMSOL ray tracing box setup with dimensions is visualized in the supplementary figure S5.

# Supplementary Information for

Rainbows in a bottle: Realizing microoptic effects by polymerizable multiple emulsion particle design

Naresh Yandrapalli[ab*], Baris Kumru[a], Tom Robinson[b*], Markus Antonietti[a]

[a] Max Planck Institute of Colloids and Interfaces, Department of Colloid Chemistry, Am Mühlenberg 1, 14424 Potsdam, Germany

[b] Max Planck Institute of Colloids and Interfaces, Department of Theory & Bio-Systems, Am Mühlenberg 1, 14424 Potsdam, Germany

Materials

All materials were used as purchased unless noted otherwise. 1-octanol (99 %, Sigma Aldrich), phenylbis(2,4,6-trimethylbenzoyl)phosphine oxide (BAPO initiator, 97%, Sigma Aldrich), Synperonic® F 108 surfactant (Sigma Aldrich). Polydi-methylsiloxane (PDMS) and curing agent were obtained as SYLGARD® 184 silicone elastomer kit from Dow Corning. 1H,1H,2H,2H-Perfluorodecyltrichlorosilane was purchased from abcr GmbH. Poly(diallyldimethylammonium chloride (PDADMAC) and poly(sodium 4-styrenesulfonate (PSS) were obtained from Sigma Aldrich. SU8 2025 (Microchem Inc.), Silicon wafer (Siegert Wafers), SU8 developer solution (Microchem Inc.) Styrene (99 %, Sigma Aldrich) was passed through alumina column to remove inhibitor before use.

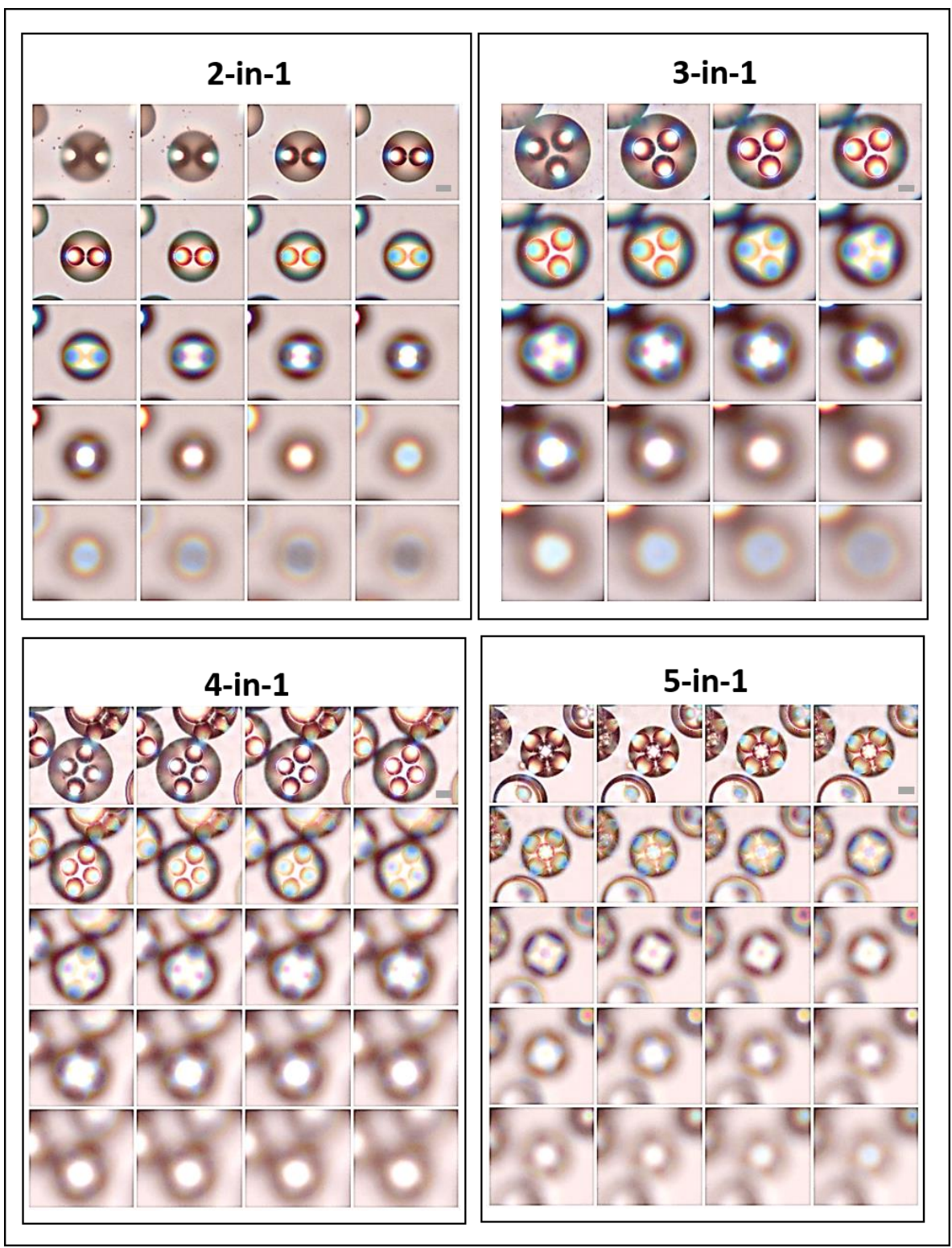

Figure S1: Bright-field z-stack of the spectral iridescence from 2-in-1, 3-in-1, 4-in-1, and 5-in-1 multi-core emulsions with each slice separated by 10 µm.

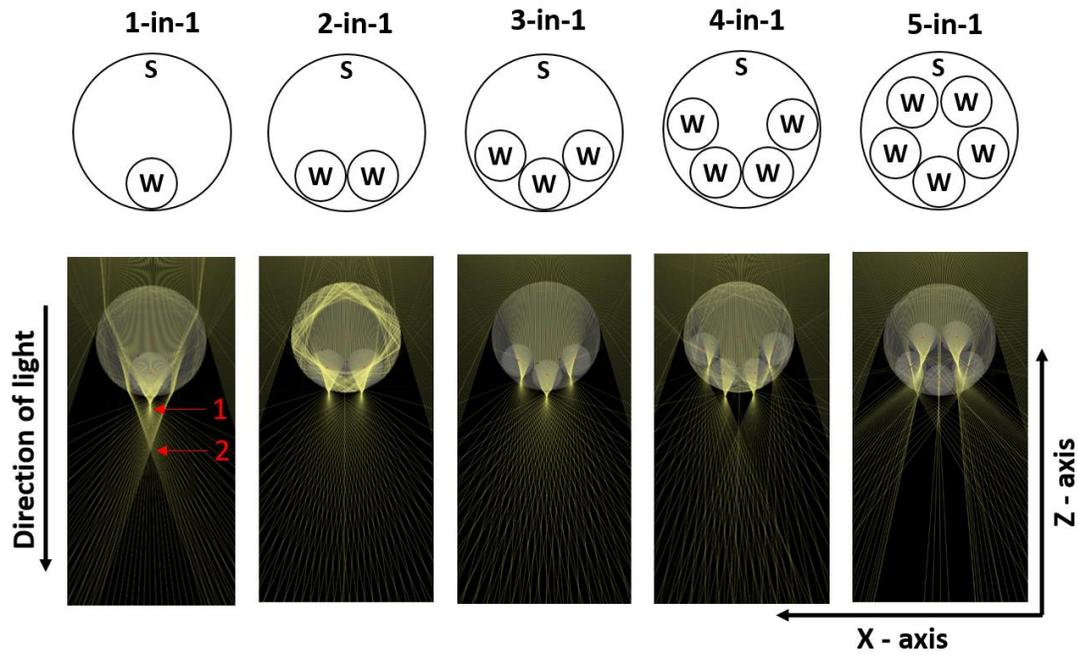

Figure S2: Comparative 2D ray-tracing of multiple emulsions. Ray tracing data reveals the formation of multiple focal points from single light source. 1 represents primary focusing point and 2- secondary focusing point. S – Corresponds to Styrene pphase and W – corresponds to aqueous phase.

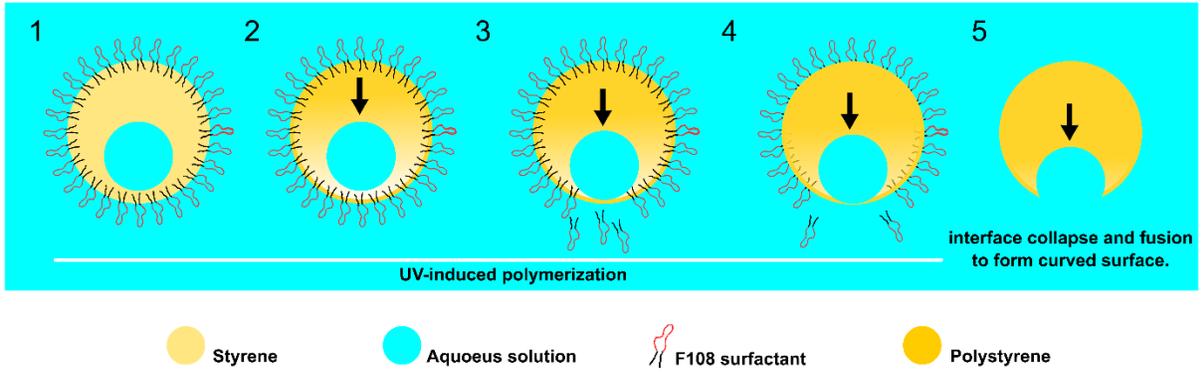

Figure S3: Mechanism for the formation of curved microspheres.

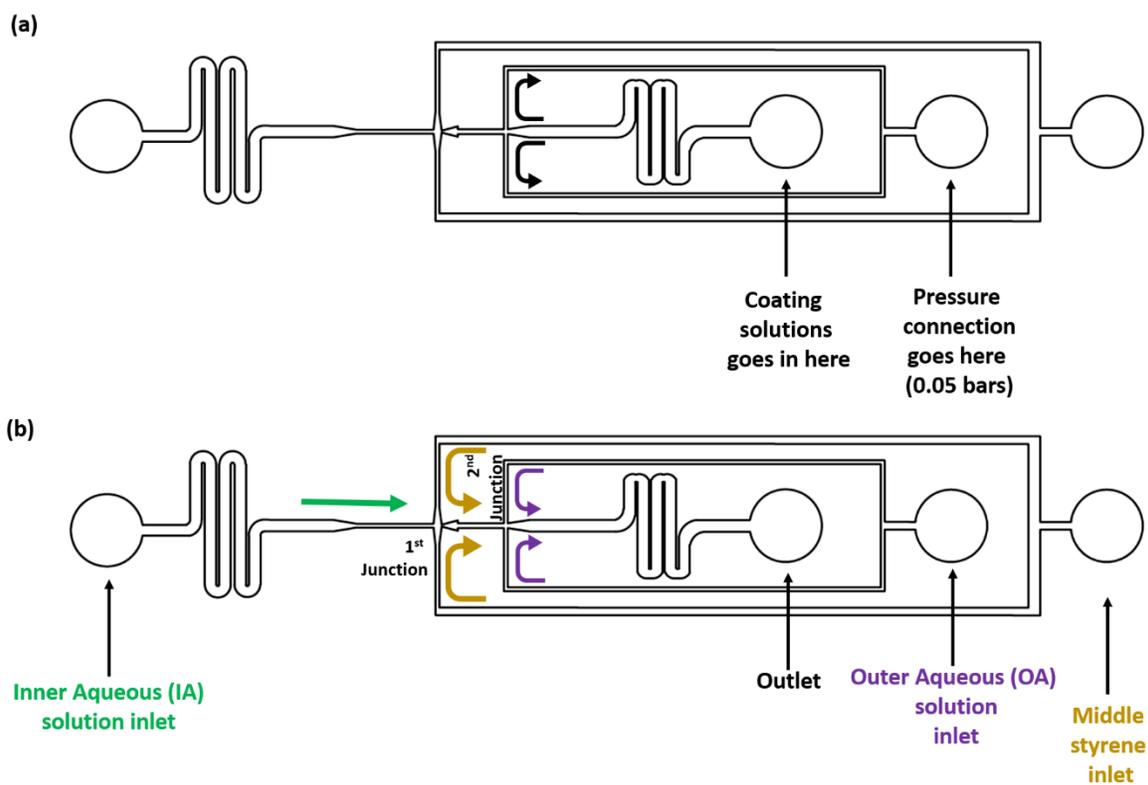

Figure S4: Microfluidic chip design with flow directions. (a) Inlet and outlets for successful surface passivation of the microfluidic chip with solutions flowing from outlet to the OA solution inlet and (b) showing multiple inlets and outlet with respective solutions and their flow direction to produce emulsions.

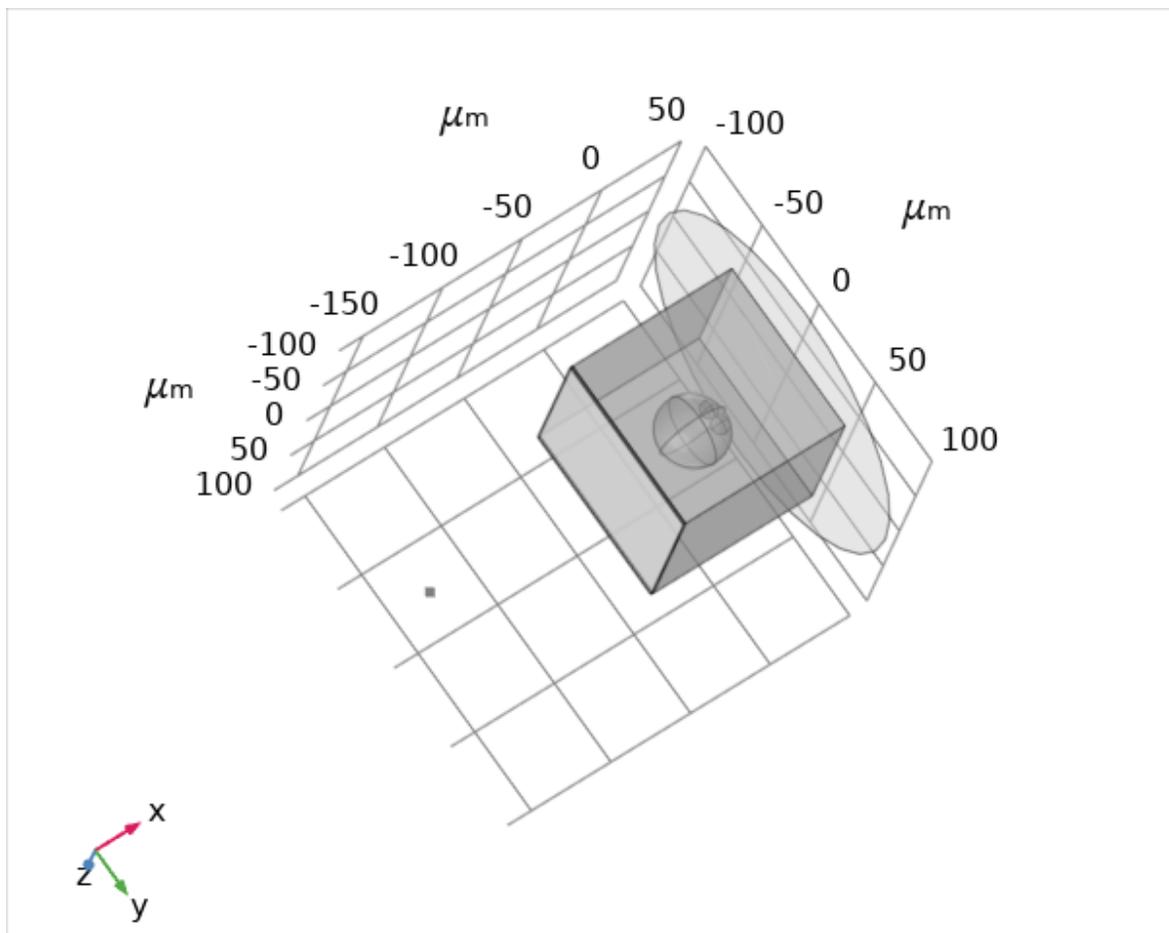

Figure S5: 3D COMSOL raytracing setup with example double emulsion with two inner cores (2-in-1) surrounded by water box topped with a glass coverslip toward the direction of the point light source positioned at (-165,0,0).

Supplementary Video 1

Z-stack video of 1-in-1 W/O/W double emulsion produced using a microfluidic device. Scale bar corresponds to 10 µm.

Supplementary Video 2

Comsol Multiphysics® ray-tracing simulation of ray-path (1000 rays) and dispersion of light along 1-in-1 W/O/W double emulsion. Colour legend depicts the wavelength of dispersed light.

Supplementary Video 3

Microfluidic production of 2-in-1 (IA - 66 mbar, Styrene – 102 mbar, OA – 112 mbar), 3-in-1 (IA - 66 mbar, Styrene – 106 mbar, OA – 105 mbar), 4-in-1 (IA - 66 mbar, Styrene – 104 mbar, OA – 94 mbar) and 5-in-1 (IA - 66 mbar, Styrene – 105 mbar, OA – 86 mbar) multi-emulsions containing aqueous inner and outer solution and middle styrene solution. Image sequence was acquired using high speed camera at ~3000 fps.

Supplementary Video 4

Z-stack video of 2-in-1 W/O/W double emulsion produced using a microfluidic device. Scale bar corresponds to 10 µm.

Supplementary Video 5

Z-stack video of 3-in-1 & 4-in-1 W/O/W double emulsion produced using a microfluidic device. Scale bar corresponds to 10 µm.

Supplementary Video 6

Z-stack video of 5-in-1 W/O/W double emulsion produced using a microfluidic device. Scale bar corresponds to 10 µm.